\documentclass[twocolumn,showpacs,preprintnumbers,amsmath,amssymb]{revtex4}

\begin{document}
\title{Quantum Transport and Integrability of the Anderson Model 
       for a Quantum Dot with Multiple Leads
      }
\author{Sam Young Cho$^1$,  Huan-Qiang Zhou$^2$, and Ross H. McKenzie$^{1,2}$}

\affiliation{$^1$Department of Physics, The University of Queensland,
             4072, Australia}
\affiliation{$^2$Centre for Mathematical Physics,
         The University of Queensland, 4072, Australia}
\date{\today}

\begin{abstract}
 We show that an Anderson Hamiltonian describing
 a quantum dot connected to multiple leads is integrable.
 A general expression for the non-linear conductance is obtained
 by combining the Bethe ansatz exact solution with 
 Landauer-B\"uttiker theory.
 In the Kondo regime, a closed form expression
 is given for the matrix conductance at zero temperature
 and when all the leads are close to the symmetric point.
 A bias-induced splitting of the Kondo resonance 
 is possible for three or more leads.
 Specifically, for $N$ leads, with each at a different chemical
 potential, there can be $N-1$ Kondo peaks in the conductance.
\end{abstract}
\maketitle
 
 {\it Introduction.}
 Since the first prediction \cite{Theory88} and 
 realization of Kondo physics in a quantum dot (QD) \cite{Exp98},
 nonequilibrium effects on the Kondo resonance due to a finite bias voltage
 across the dot
 have attracted increasing attention.
 In the experiments, the zero bias peak of the differential conductances
 has been observed as a signature
 of the Kondo effect on electron transport through a QD.
 In the unitary scattering limit,
 observations of perfect transmission \cite{Wiel00,Ji02}
 provide further evidence for
 the Kondo effect in QDs.
 The nonequilibrium density of states (DOS) of the dot 
 has been predicted \cite{Meir93}
 to exhibit a splitting of the Kondo peak
 due to a bias voltage applied between the source and the drain.
 This splitting has not been observed in transport measurements.
 To observe the splitting of the Kondo resonance by a finite
 voltage bias,
 an experiment with extra leads \cite{Sun01,Lebanon02} 
 has been proposed.
 Very recently, 
 such a splitting 
 was observed in an experiment \cite{Franceschi02}
 where a three-lead setup was employed.

 In a conventional bulk Kondo system 
 \cite{Hewson} (e.g., a magnetic impurity in a metal),
 there is a single chemical potential and
 the Kondo resonance in the DOS appears at the Fermi energy
 due to the formation of a singlet 
 between the local moments of the impurity 
 and the conduction electrons. 
 If the impurity has available a second
 conduction band to form singlet
 states, a second Kondo resonance in the DOS 
 might be expected to occur
 at the chemical potential of the second conduction band.
 The splitting of the Kondo resonance of a QD by 
 the differential chemical potentials of the two leads 
 then seems to be reasonable.
 However, it is not still clear why 
 the differential conductance has only a single peak at zero bias 
 in experiments with two leads.
 Thus there arises a fundamental question associated with a Kondo resonance
 in a system with several chemical potentials
 that can be fabricated in nano-scale electronic devices:
 why the split Kondo peaks have not been seen in two-lead systems?
 To help answer this question we consider a QD coupled to 
 multiple leads.
 The QD is described by
 an Anderson model generalized to a multiple-lead one.
 It will be shown that the multiple-lead Anderson model is integrable
 and exactly solvable by a unitary transformation
 and the Bethe ansatz \cite{Kawakami,Wiegmann,Filyov,Tsvelick83}.
 By using the exact solution,
 a general expression for the conductance of the $N$-lead system
 shows that the Kondo resonance at equilibrium 
 is split into $N-1$ peaks
 by increasing the difference between the chemical potentials
 of the different leads.
 This then clearly shows why only a single peak of the conductance 
 occurs in the two-lead system.

 {\it Model.}
 We consider an Anderson model in which $N$ leads are coupled to
 the QD, as in Fig. 1.
 The leads are described under the {\it unfolded} formalism with fermions.
 Within this formalism, 
 fermions incident on the dot $(x=0)$ from a lead $( x < 0)$ 
 are scattered away from the dot to leads $( x > 0)$.
 In the continuum limit, the multiple-lead Anderson model
 Hamiltonian is given by
 \begin{eqnarray}
  H \!\!&=&\!\!-i \sum^{N}_{m=1;\sigma} \int^{\infty}_{-\infty} dx 
      \; c^\dagger_{m\sigma}(x) \partial_x c_{m\sigma}(x)
   + \sum_{\sigma} \varepsilon_d d^\dagger_\sigma d_\sigma 
   \nonumber \\ && \hspace*{1.1cm}
 +U n_\uparrow n_\downarrow
 +\sum^{N}_{m=1;\sigma}V_m 
    ( c^\dagger_{m\sigma}(0) d_\sigma\!\!+\!{\rm h.c.}),
 \end{eqnarray}
 where $n_\sigma = d^\dagger_\sigma d_\sigma$
 is the number of electrons of spin $\sigma$ on the dot
 and $U$ is the onsite Coulomb repulsion.  
 $c_{m\sigma}$ and $d_\sigma$ are the annihilation operators
 with spin $\sigma$ for electrons in the lead $m$ and
 the dot. $\sum_m$ is a sum over the multiple leads ($m = 1, \cdots, N$).  
 $\varepsilon_d$ is the energy level on the dot.
 Here the hopping amplitudes between the dot
 and the lead $m$, $V_m$, are allowed to be arbitrary.

 Previously, it has been shown that,
 for the $N=2$ case, a unitary (Bogoliubov) transformation
 can be used to transform the Hamiltonian
 to a single-lead Anderson Hamiltonian \cite{Konik01}.
 We now generalize this to the case of general $N$.
 To do this, one performs 
 a unitary transformation,
 $ {\widetilde {\bf c}} = {\bf U}_N\ {\bf c}$,
 for the lead electrons,
 where $ {\bf c}= ( c_1, \cdots, c_{N}) $
 and 
 ${\widetilde {\bf c} } = ( {\widetilde c}_1, \cdots, {\widetilde c}_{N})$.
 The components of the $N \times N$ matrix ${\bf U}_N$ are 
 a function of the hopping amplitudes, $V_m$.
 ${\bf U}_N$
 should satisfy ${\bf U}^{\dagger}_N{\bf U}_N=I$.
 If 
 (i)  $\sum_{m} V_m [{\bf U}_N]_{mm'}=\sum_{m} [{\bf U}^\dagger_N]_{m'm}V_m$,
 and 
 (ii) $\sum_{m} V_m [{\bf U}_N]_{mm'}=\sqrt{\Gamma}$ for $m' = 1$
 and $0$ for $m' \neq 1$, 
 one obtains the one-lead Anderson Hamiltonian
 and $N-1$ free fermion Hamiltonians.
 Then a $N\times N$ unitary matrix for the multiple leads 
 has a form satisfying with $[{\bf U}_N]_{1m}=V_m/\Gamma$
 and $\Gamma=\sum_{m} V^2_m$.
 For $N > 2$, actually, there are more freedoms to choose
 a unitary matrix. The freedoms give us different matrices 
 for a unitary
 transformation acting only on 
 $({\tilde c}_2, \cdots, {\tilde c}_N)$, but leaving ${\tilde c}_1$ invariant, 
 which does not affect the physics.

 As a consequence, the unitary transformation satisfying such conditions
 decomposes the multiple-lead Hamiltonian
 into $N$ independent sub-Hamiltonians, ${\widetilde H}_m$, as
 \begin{equation}
  H = \sum_m {\widetilde H}_m,
 \end{equation}
 where
 \begin{eqnarray}
  {\widetilde H}_1
  \!\!\! &=&\!\!\sum_{\sigma}\left[ -i \int^{\infty}_{-\infty} \; dx \; 
    {\widetilde c}^\dagger_{1\sigma}(x) \partial_x {\widetilde c}_{1\sigma}(x)
   + \varepsilon_d d^\dagger_\sigma d_\sigma 
   \right.  \nonumber \\ && \left. \hspace*{1.7cm}
   + U n_\uparrow n_\downarrow
  \! +\! \sqrt{\Gamma}
    ( {\widetilde c}^\dagger_{1\sigma}(0) d_\sigma + {\rm h.c.} ) \right]\! ,
   \\ 
  {\widetilde H}_{m}
  \!\!\! &=&\!\!-i \! \sum_{\sigma}\!\! \int^{\infty}_{-\infty} \!\! dx \;
    {\widetilde c}^\dagger_{m\sigma}(x) 
    \partial_x {\widetilde c}_{m\sigma}(x)
  \mbox{~~for $ m\! \in \! [2,N]$}.
 \end{eqnarray}
 This is a generalization of the $N=2$ case treated in
 Ref. \cite{Konik01}.
 The transformed Hamiltonian can be solved exactly
 because the sub-Hamiltonian, ${\widetilde H}_1$,
 is the one-lead Anderson model that is exactly solvable
 via the Bethe ansatz 
 \cite{Kawakami,Wiegmann,Filyov,Tsvelick83}.

\begin{figure}
\vspace*{6cm}
\includegraphics{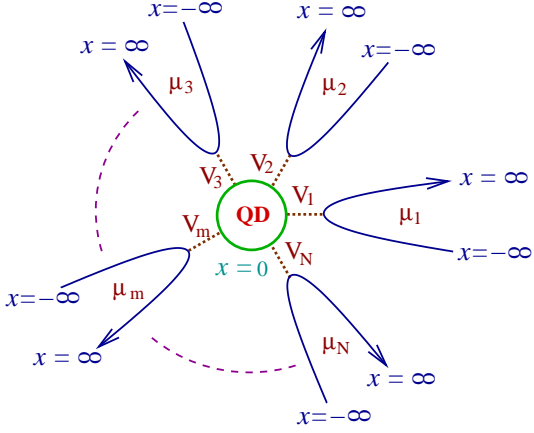}
\caption{A quantum dot (QD) coupled to $N$-multiple leads.
         $V_m$ is the tunneling amplitude between the $m$-th 
         lead and the QD. 
         $\mu_m$ is the chemical potential of the $m$-th lead.
         The leads are presented under
         the unfolded formalism.}
\end{figure}
 {\it Integrable excitations and scattering amplitudes.}
 The scattering amplitudes of electronic excitations off the QD
 coupled to the $N$ leads can be calculated based on the 
 exact solution of ${\widetilde H}_1$.
 In the transformed $N$ leads,
 the integrable excitations, $\{{\widetilde \psi}_m\}$,
 will scatter off the dot with 
 some pure phase shift
 with spin $\sigma$, $\delta^\sigma_1(\varepsilon )$, where 
 in particular $\delta^\sigma_m(\varepsilon)=0$ for $m \in [2,N]$.  
 With the unfolded formalism,
 the scattering can be described by the relation 
 \begin{equation}
  {\widetilde \psi}_m (x>0)= e^{i\delta^\sigma_m} {\widetilde \psi}_m(x<0).
 \label{excitation}
 \end{equation}
 Equation (\ref{excitation}) leads to the scattering amplitudes
 $S^\sigma_{mm'}(\varepsilon)$ of electronic excitations, $\{\psi_m\}$,
 of energy $\varepsilon$ between leads in the multiple-lead system.
 Assuming the relation $\psi_m = \sum_{mm'}[{\bf U}_N]_{mm'} 
 {\widetilde \psi}_{m'}$,
 the scattering matrix is straightforwardly given by
 \begin{eqnarray}
 S^\sigma_{mm'}(\varepsilon) = \delta_{mm'}
  + 2i \Gamma_{mm'}e^{i\frac{\delta^\sigma_1}{2}} 
    \sin \frac{\delta^\sigma_1}{2},
 \label{scattering}
 \end{eqnarray}
 where $\Gamma_{mm'} =
    [ {\bf U} {\bf P}_1 {\bf U}^{-1}]_{mm'}$ 
 and ${\bf P}$ is a polarization matrix:
 $[{\bf P}_m]_{mm} = 1$ and other entries are zero.
 For $m \neq m'$, $S^\sigma_{mm'}$ is a transmission amplitude 
 $T^\sigma_{mm'}$
 from $m'$ to $m$.
 For $m = m'$, $S^\sigma_{mm}$ corresponds to a reflection amplitude
 $R^\sigma_{mm}$
 from $m$ to $m$.
 From $\Gamma_{mm'} = \Gamma_{m'm}$, 
 $T^\sigma_{mm'}(\varepsilon)=T^\sigma_{m'm}(\varepsilon)$ is automatically
 preserved.
 

 {\it Differential matrix conductance.}
 The current and the conductance through the QD
 can be obtained by the Landauer-B\"uttiker theory \cite{Datta95}
 for quantum transport through nano-devices.
 To describe scattering away from the Fermi energy
 and calculate the differential conductance,
 we employ an ansatz \cite{Konik01} verified
 in Refs. \cite{Fendley95,Saleur00}.
 The ansatz allows us to use 
 the in-equilibrium scattering matrices 
 to calculate the contribution to the current of any given excitation.  
 Konik and coworkers discussed the details
 of the implementation of the nonequilibrium computation
 in Ref. \cite{Konik02}.
 With
 $T^\sigma_{mm'}(\varepsilon)=T^\sigma_{m'm}(\varepsilon)$,
 at zero temperature,
 the current in lead $m$ is given by 
 \begin{equation}
 I_m = \frac{e}{h} \sum_{m'\neq m;\sigma}
         \int^{\mu_{m}}_{\mu_{m'}}  d\varepsilon
      \;\; \Big|T^\sigma_{mm'}(\varepsilon,\{\mu_m\})\Big|^2 ,
 \label{current}
 \end{equation}
 where $\mu_m$ is the chemical potential at the lead $m$ and
 \begin{equation}
   \Big|T^\sigma_{mm'}(\varepsilon,\{\mu_m\})\Big|^2 
  = 4 \Gamma^2_{mm'}
    \sin^2\left[\frac{1}{2} \delta^\sigma_1(\varepsilon,
     \{\mu_m\})\right].
 \label{transmission}
 \end{equation}
 
 To determine $\delta_1$, we solve ${\widetilde H}_1$ via the Bethe ansatz 
 for the one-lead Anderson model. 
 The integrability of ${\widetilde H}_1$ leads to a set of
 quantization conditions identical to that of the one-lead Anderson model.
 Single particle excitations with momenta $\{k_j\}$ are identified
 by an appropriate basis.
 Scattered particle eigenstates from the dot picks up
 the {\it bare} phase $\delta(k)=-2\tan^{-1}[\Gamma/(k-\varepsilon_d)]$.
 Calculating two particle eigenstates makes it possible to 
 get the scattering matrices of excitations.
 The scattering matrices satisfying a Yang-Baxter relationship 
 are identical to that of the one-lead Anderson model.
 Then a set of $N_e$ multi-particle eigenstates carrying total spin
 $S_z=N_e/2-M$ should
 satisfy the quantization conditions \cite{Kawakami,Wiegmann,Filyov} as
 \begin{eqnarray}
 e^{ik_j L + i\delta(k_j)} &=&
 \prod^M_{\alpha=1} 
 \frac{g(k_j)-\lambda_\alpha+i/2}{g(k_j)-\lambda_\alpha-i/2} , \nonumber \\
 \prod^M_{\beta=1} 
 \frac{\lambda_\alpha-\lambda_\beta+i}{\lambda_\alpha-\lambda_\beta-i}
  &=&
 -\prod^{N_e}_{j=1} 
 \frac{g(k_j)-\lambda_\alpha-i/2}{g(k_j)-\lambda_\alpha+i/2},
 \end{eqnarray}
 where $g(k)=(k-\varepsilon_d-U/2)^2/2U\Gamma$
 and $M$ characterizes the spin projection of the system
 with the auxiliary parameters, $\{\lambda_\alpha\}$.
 For $\varepsilon_d > -U/2$, then,
 $N_e$ total momenta $k$'s form an $N_e$ particle ground state configuration.
 $N_e-2M$ of $N_e$ momenta $k$'s is real and $2M$ is complex
 via $M$ real $\lambda_\alpha$'s.
 The $2M$ complex momenta are given by
 $k^{\pm}_\alpha=x(\lambda_\alpha)\pm i y(\lambda_\alpha)$ with
 $x(\lambda)=U/2+\varepsilon_d-\sqrt{U\Gamma}
 [\lambda+(\lambda^2+1/4)^{1/2}]^{1/2}$
 and
 $y(\lambda)=\sqrt{U\Gamma} [-\lambda+(\lambda^2+1/4)^{1/2}]^{1/2}$.

 According to Andrei's procedure for determining the momentum, $p$, 
 of an added electron in a periodic system of size $L$ \cite{Andrei82},
 the quantization condition of the system leads to $p = 2\pi n/L$.
 Contributions to the momentum come from the bulk of the system
 and the dot:
 $$ p = 2\pi n /L=p_{\rm bulk}+p_d/L.$$
 The dot contribution scaled by the size of the system
 is identified with the scattering phase of the excitation off the dot,
 which gives the relation between the phase and the momentum from the dot
 as $\delta_1 = p_d$.
 In adding an electron with spin $\sigma$ to the system, then,
 the electron scattering phase shift has two contributions from the
 charge, $p^{Q}$, and the spin sectors, $p^{S}$,  \cite{Konik01} as given by
 \begin{equation}
  \delta^\sigma_1 = p^\sigma_{d} = p^{Q}_{d} (k) + p^{S}_{d} (\lambda).
  \label{phase}
 \end{equation}
 The electronic scattering phase shifts are related to the density
 of states $\rho_d(k)$ and $\sigma_d(\lambda)$ by the equations: 
 \begin{eqnarray}
 p^Q_d(k)\! &=&\! \delta(k) \!+\! \int^{\tilde q}_q \! d\lambda 
             [ \theta_1(g(k)\!-\!\lambda)\! -\! 2\pi ] \sigma_d(\lambda),
  \\
 p^S_d(k)\! &=&\! {\tilde \delta}(k) \!+\! \int^{\tilde q}_q \! d\lambda' 
             [ \theta_2(g(k)\!-\!\lambda')\! -\! 2\pi ] \sigma_d(\lambda')
 \nonumber \\ && \hspace*{0.6cm}
             +\! \int^{B}_{-D} \! dk 
             [ \theta_1(\lambda\!-\!g(k))\! -\! 2\pi ]\  \rho_d(k),
 \end{eqnarray}
 where ${\tilde \delta}=2 {\rm Re}[\delta(x(\lambda)+iy(\lambda)]$.
 $q/B$ are the Fermi surfaces of the seas of $k$ and $\lambda$
 excitations while ${\tilde q}$ is related to the band cutoff, $D$.
 Here $\theta_{1,2}$ for describing the dot momentum should be chosen
 to ensure that
 $p^Q_d(k \rightarrow -\infty)=p^S_d(\lambda \rightarrow \infty)=0$.
 Moreover, the dot momenta are simply related to the dot density
 of states:
 \begin{eqnarray}
 \partial_k p^Q_{d}(k) = 2 \pi \rho_{d} (k), 
 \mbox{~and~} 
 \partial_\lambda p^S_{d}(\lambda) = -2 \pi \sigma_{d} (\lambda).
 \end{eqnarray}
 Integrating the density of states gives us the dot momenta.  
 Consequently, the scattering phase shift is given by
 \begin{equation}
 \delta^\sigma_1 = 2\pi \int^B_{-D} dk \rho_d(k) 
                 + 2\pi \int^{\tilde q}_{q} d\lambda' \sigma_d(\lambda').
 \end{equation}
 This phase shift satisfies 
 the Langreth-Friedel sum rule, $\delta^\sigma_1=2\pi n_\sigma$,
 relating the phase shift to
 the total number of electrons $n_d$ in the dot
 \cite{Langreth66}. 

 To obtain the matrix conductance of the multiple-lead system
 away from the symmetric point $(\varepsilon_d-\mu_m = -U/2)$,
 we need to do a numerical calculation for the associated integral equations.
 But at the symmetric point
 the scattering phase shift is obtained by using 
 an exact expression for $\rho_d(k<0)$ \cite{Tsvelick83}
 and a direct relation between
 the phase shifts for the electron with spin $-\sigma$
 and the hole with spin $\sigma$
 from a property of electron-hole transformation 
 based on the SU(2) spin symmetry.
 The phase shift is given by \cite{Konik02}
 \begin{equation}
 \delta_1(\varepsilon)\!
 = \!\frac{3}{2}\pi \!
  -\! \sin^{-1}\!\!\left[
    \frac{4T_{K,m}^2-\pi^2(\varepsilon-\mu_m)^2}
         {4T_{K,m}^2+\pi^2(\varepsilon-\mu_m)^2}
            \right]\!+ C(\varepsilon),
 \label{phase2}
 \end{equation}
 where
 the Kondo temperature for a lead at chemical potential $\mu_m$ is
 $$T_{K,m} \!\!
  =\!\! \sqrt{\frac{U\Gamma}{2}} 
    \exp\!\!\left[\frac{\pi}{2\Gamma U}
     \left[(\varepsilon_d\!-\!\mu_{m})(\varepsilon_d\!-\!\mu_{m}+U)
     \! - \!\Gamma^2\right]
        \right].$$
 Here, $C(\varepsilon)$ does not give any significant phase shift 
 when the Kondo energy scale is much smaller than the Coulomb
 interaction $U$.
 For $|\mu_m-\mu_{m'}| \ll U$,
 we can assume all of the leads are at the symmetric point.
 This makes it possible to take into account the essence
 of the physics associated with the splitting of the Kondo resonance
 in a multiple-lead system.
 Then one can obtain a simple expression for the matrix conductance
 ($G_{mm'}=-e\partial_{\mu_{m'}}I_m$) 
 from Eq. (\ref{current}), (\ref{transmission}) and (\ref{phase2}).
 The matrix conductance in the multiple-lead Kondo-dot system
 is given by
 \begin{eqnarray}
 G_{mm} \!\!\! &= &\! -\sum_{m' \neq m} G_{mm'} ,
 \label{conductance1}
  \\
 G_{mm'\atop (m \neq m')}\!\!\!\!\!  &= &\!\!\! - 4 G_0
     \Gamma^2_{mm'} \!\!
     \left[\! 1\! +\! \frac{\pi^2}{4}\!\! \left(
       \frac{\mu_m-\mu_{m'}}{T_{K,{\rm max}[\mu_m,\mu_{m'}]}}
    \!\right)^{\! 2}\!\right]^{-1} \!\!\!\!\!\!\! ,
 \label{conductance2}
 \end{eqnarray}
 where 
 $G_0=2e^2/h$ is the quantum of conductance, and
 $\Gamma_{mm'} = V_m V_{m'}/\Gamma$.
 This multiple-lead matrix conductance 
 is the generalized expression of  the conductance 
 for the two-lead Kondo-dot system.
 It reduces to the conductance in the two-lead system \cite{Konik02}.
 For a symmetric coupling ($V_1=\cdots=V_N$)
 and $\mu_1 = \cdots = \mu_N$, 
 the matrix conductance is 
 $G_{mm}/G_0=4(N-1)/N^2$ and $G_{mm'}/G_0=-(2/N)^2$.
 The resultant matrix conductance agrees with that of a multi-lead 
 quantum point-contact 
 for free fermions \cite{Nayak99}.
 This unitary scattering limit
 shows the Fermi liquid nature of the multiple-lead Kondo-dot system.

 Note that the multiple-lead matrix conductance 
 in Eq. (\ref{conductance1}) and (\ref{conductance2}) shows clearly
 that a conductance peak for the transmission from $m$ to $m'$ 
 is developed 
 when the two chemical potentials are tuned to be equal, $\mu_m=\mu_{m'}$.
 As the chemical potential difference increases,
 the amplitude of the conductance decreases.
 In a $N$-lead system, if every chemical potential has 
 a different value, 
 the conductance $G_{mm}$ versus $\mu_m$ has a total of  
 the $N-1$ conductance peaks, one at each of the
 other chemical potentials. 
 The amplitude of the conductance $G_{mm'}$ versus $\mu_m$ has its 
 maximum value for $\mu_m=\mu_{m'}$.
 The maximum values of $G_{mm'}$'s
 have a one-to-one correspondence to 
 the conductance peaks of $G_{mm}$.
 This behavior of the conductances implies 
 that electrons from each lead 
 participate in screening the local moment of the dot
 and take part in forming a single Kondo resonance at equilibrium. 
 Increasing the difference between the chemical potentials,
 the electrons from each of the $N$ leads 
 have their own Kondo resonances with the dot.
 Each resonance is 
 characterized by 
 a Kondo temperature, $T_{K,m}$, 
 depending on the value of the chemical potential of the lead.
 Since each lead creates a single lead-dot Kondo resonance,
 the $N$-lead system has $N$ lead-dot Kondo resonances.
 If the chemical potentials of two of
 the leads are adjusted to be equal
 then the two Kondo resonances corresponding to these leads
 merge together in $G_{mm'}$. Then this results in
 only a single transmission peak in the conductance $G_{mm}$.
 Therefore,
 an electron transport measurement in the two-lead system
 is able to capture 
 only the single transmission peak 
 even though there are two lead-dot Kondo resonances 
 created by the two leads. 
 Hence,
 the two-lead system is not a good probe to observe
 the splitting of the Kondo resonance by finite biases. 
 
 {\it Three-lead and four-lead system.}
 Before proceeding to the conclusion,
 we discuss the conductance for the three leads $(N=3)$
 and the four leads $(N=4)$.
 The unitary transformation for the three-lead system
 is given by the unitary matrix; 
 \begin{equation}
 {\bf U}_{3} = \frac{1}{\sqrt{\Gamma}} 
             \left(\begin{array}{ccc}
              V_1 & V_2 & V_3 \\
              V_2 & a  & b \\
              V_3 & b  & c
             \end{array}\right),
 \end{equation}
 where
 $a=(-V_1V^2_2+V^2_3\sqrt{\Gamma})/\gamma$, 
 $b=(-V_1V_2V_3-V_2V_3\sqrt{\Gamma})/\gamma$, and 
 $c=(-V_1V^2_3+V^2_2\sqrt{\Gamma})/\gamma$ 
 with ${\gamma}=V^2_2+V^2_3$. 
 It can be obtained explicitly under the necessary condition 
 we discussed above.
 Similarly, the unitary matrix ${\bf U}_4$ for four leads can be determined.

\begin{figure}
 \vspace*{5.7cm}
 \includegraphics{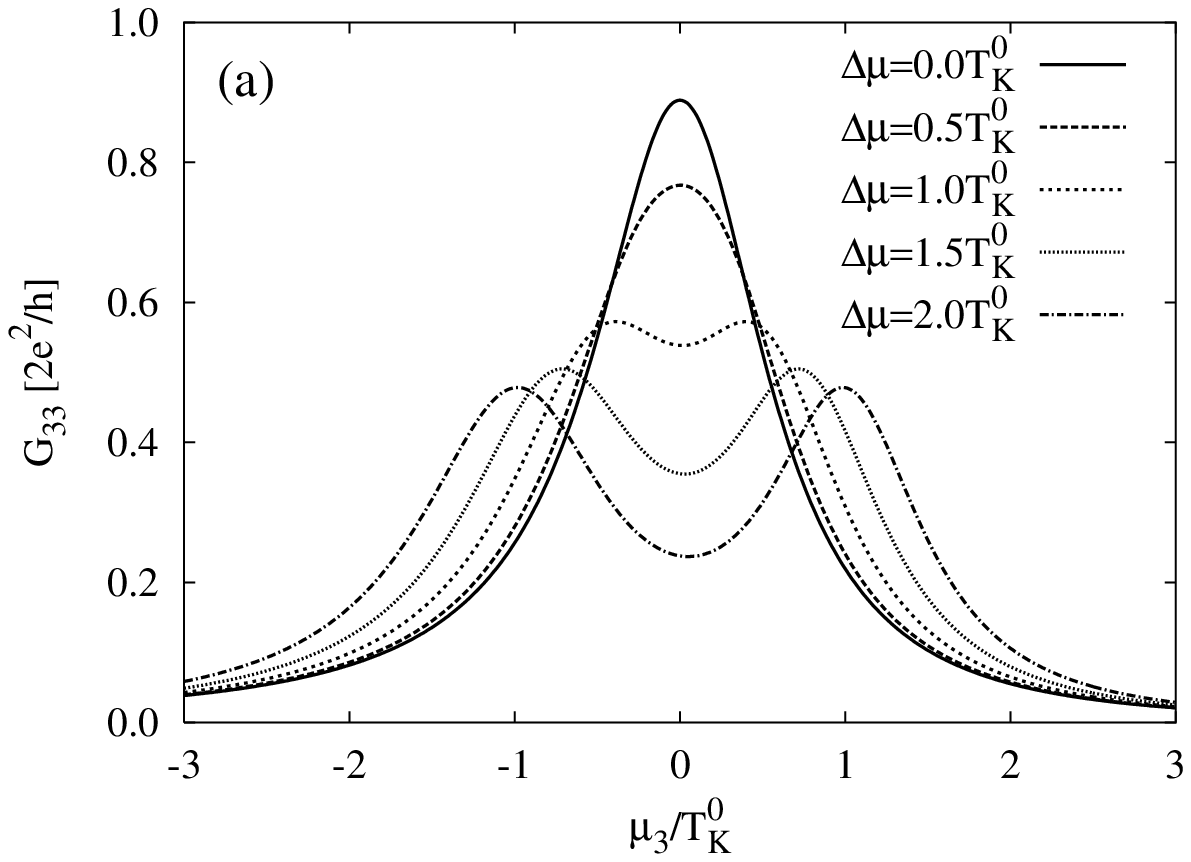}
 \vspace*{5.5cm}
 \includegraphics{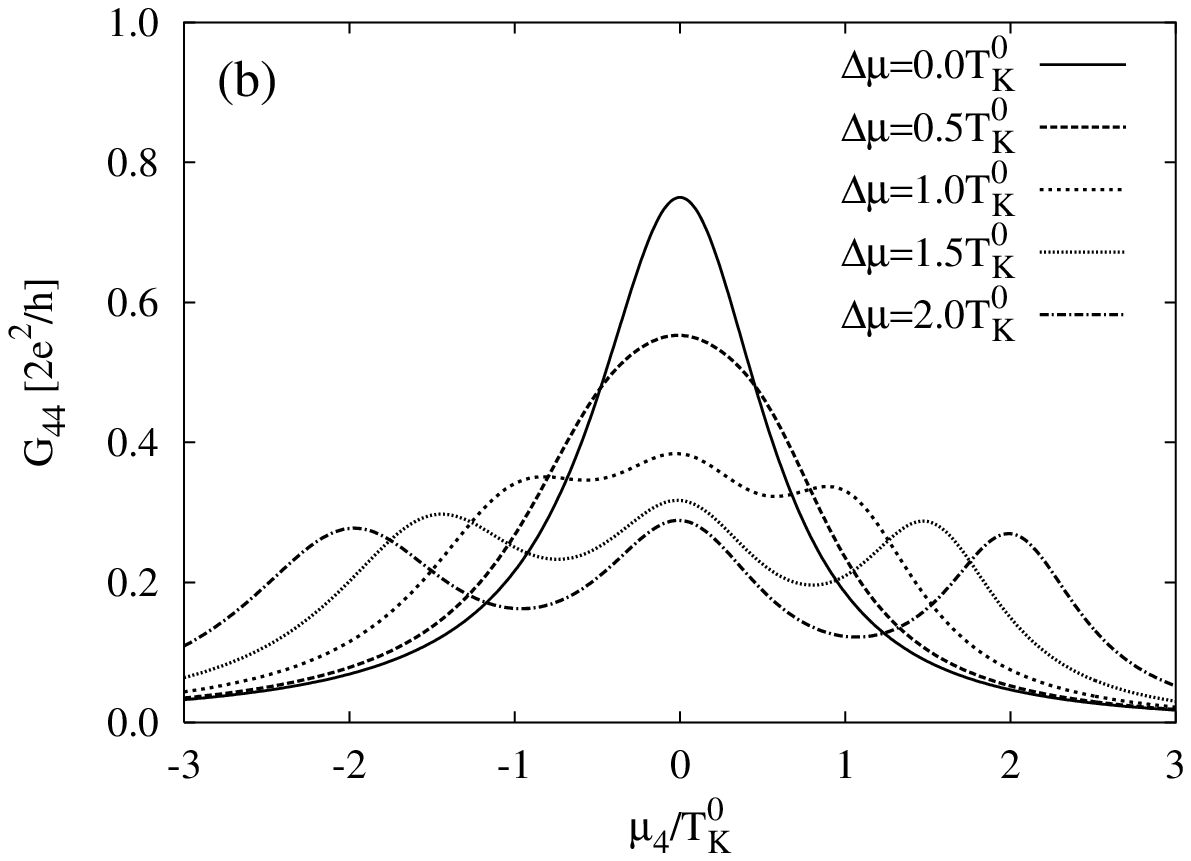}
 \caption
   {Splitting of the Kondo resonance by multiple leads.
    (a)  Conductance $G_{33}$ as a function of the chemical potential $\mu_3$
         for a quantum dot symmetrically coupled
         (i.e., $V_1=V_2=V_3$) to the three leads ($N=3$),
          $\Delta\mu=\mu_2-\mu_1$.
    (b)   Conductance $G_{44}$ as a function of the chemical potential $\mu_4$
          for a quantum dot  symmetrically coupled to the four leads ($N=4$), 
           $\Delta\mu=\mu_2-\mu_1=\mu_3-\mu_2$.
          The temperature is zero,
          $U=100.0\Gamma$, and $\varepsilon_d=-3.0\Gamma$. 
          $T^0_K$ is the Kondo temperature at equilibrium. 
          Here, at equilibrium, all chemical potentials are set to be zero.
          }
 \label{fig2}
\end{figure}

 We plot the conductance $G_{33}$ as a function of $\mu_3$ 
 for $N=3$
 and the conductance $G_{44}$ as a function of $\mu_4$ for $N=4$
 in Fig. \ref{fig2} (a) and (b), respectively. 
 When all the leads are at the same chemical potential ($\Delta\mu=0$), 
 the amplitude of the conductance is shown to be reduced
 as the number of leads increases. 
 The maximum amplitudes are $G_{33}/G_0=8/9$ and $G_{44}/G_0=3/4$.
 As the difference between the other chemical potentials, $\Delta\mu$, 
 become larger than the Kondo temperature $T^0_K$ at equilibrium,
 the single peak at $\Delta\mu=0$ splits progressively 
 into two and three peaks for three and four leads, respectively.
 Figure \ref{fig2} (a) shows that
 for $\Delta\mu \simeq 2 T^0_K$, the amplitudes of the split peaks
 reduce to around half the value of that of the equilibrium Kondo
 peak ($\Delta\mu=0$).
 The suppression of the Kondo resonance is on a voltage scale $T^0_K$.  
 This behavior agree qualitatively,
 but not quantitatively, with the experimental 
 results in Ref. \cite{Franceschi02}.

%

 {\it Summary.}
 By using a unitary transformation and the Bethe ansatz, 
 the multiple-lead Anderson model is shown to be integrable. 
 A general expression for  the matrix conductance from the integrability
 has been obtained.
 The conductance for the $N$-lead system
 shows $N-1$ split Kondo peaks located 
 at $N-1$ different chemical potentials.
 This 
 shows that a Kondo-dot system with multiple leads provides 
 a good probe to observe the nonequilibrium effects on the Kondo
 resonance by a voltage bias in transport measurement.

 {\it Acknowledgments.}
 This work was supported by the Australian Research Council.


\end{document}